# SPIN FRUSTRATION AND MAGNETIC EXCHANGE IN COBALT ALUMINUM OXIDE SPINELS

N. Tristan,[1] V. Zestrea,[1,2] G. Behr,[1] R. Klingeler,[1] B. Büchner, [1]
H.A. Krug von Nidda,[3] A. Loidl,[3] and V. Tsurkan[2,3*]

[1] *Leibniz Institute for Solid State and Materials Research, IFW- Dresden,
D-01171, Dresden, Germany*
[2]*Institute of Applied Physics, Academy of Sciences of Moldova, MD-2028, Chisinau,
R. Moldova*
[3]*Experimental Physics V, Center for Electronic Correlations and Magnetism, University of
Augsburg, D 86159, Augsburg, Germany*

(Dated: November 5, 2007)

## Abstract

We report on x-ray diffraction, magnetic susceptibility, electron–spin resonance and heat-capacity studies of $Co[Al_{1-x}Co_x]_2O_4$ for Co concentrations $0 \leq x \leq 1$. In this spinel system only the A-site $Co^{2+}$ cation is magnetic, while the non-magnetic B-site $Al^{3+}$ is substituted by the low-spin non-magnetic $Co^{3+}$, and it is possible to investigate the complete phase diagram from $Co^{2+}Al^{3+}{}_2O_4$ to $Co^{2+}Co^{3+}{}_2O_4$. All samples reveal large negative Curie-Weiss temperatures $\Theta_{CW}$ of the order of -110 K independent of concentration, which is attributed to a high multiplicity of the superexchange interactions between the A-site $Co^{2+}$ cations. A pure antiferromagnetic state is found for $x = 1.0$ and 0.9 with Néel temperatures $T_N = 29.5$ K and 21.2 K, respectively, as evidenced by lambda-like anomalies in the specific heat. Compositions with $0.3 \leq x \leq 0.75$ show smeared out strongly reduced magnetic ordering temperatures. At low temperatures, a $T^{2.5}$ dependence of the specific heat is indicative of a spin-liquid state. For $x \leq 0.2$ a $T^2$ dependence of the specific heat and a spin-glass like behavior of the susceptibility below $T_f \sim 4.7$ K are observed. The high value of the frustration parameter $f = |\Theta_{CW}|/T_f > 10$ indicates the presence of strong spin frustration at least for $x \leq 0.6$. The frustration mechanism is attributed to competing nearest neighbor and next-nearest neighbor superexchange interactions between the A-site $Co^{2+}$ ions.

---

* Corresponding author, e-mail:Vladimir.Tsurkan@physik.uni-augsburg.de

## I. INTRODUCTION

Correlated electron systems with frustrated magnetic exchange are subject of intensive current theoretical and experimental studies.[1] The inability to satisfy simultaneously all pair-wise interactions between spins results either in a spin-glass state for structurally disordered magnets[2] or in a dynamic liquid ground state[3] in magnets with topological constraints yielding exotic ordered phases, like, for example, spin-ice.[4,5]

Magnetic $AB_2X_4$ spinels recently have attracted much attention because of strong frustration effects. Magnetic ions on the B-sites of the spinel lattice form a network of corner-sharing tetrahedra, an analog of the pyrochlor lattice, which is known as the most frustrated magnetic lattice.[6] Antiferromagnetic (AFM) spinels with the magnetic ions only on the B-sites reveal unconventional magnetic ground states. Lee *et al.*[7] reported on composite spin degrees of freedom that emerge from frustrated interactions in cubic $ZnCr_2O_4$. Ueda *et al.*[8] observed a metamagnetic transition to one half magnetization plateau in frustrated $CdCr_2O_4$. The direct Cr-Cr AFM exchange dominates in chromium oxide spinels and the resulting geometric frustration is released by a Peierls-like structural transition attributed to a spin-driven Jahn-Teller effect.[9-11] Similar structural anomalies which accompany magnetic transitions were also observed in the AFM chromium sulfide and selenide spinels.[12-14] They were ascribed to another type of frustration, called bond frustration, related to competing antiferromagnetic and ferromagnetic superexchange interactions which equally dominate the magnetic exchange in chromium chalcogenides. Due to geometrical and bond frustration the Cr spinels reveal a rich phase diagram,[15] that recently has been explained by band-structure calculations including a modeling of the effective exchange constants between the Cr spins.[16]

Strong frustration effects were recently found in a number of sulfide and oxide spinels containing magnetic ions solely on the A-sites.[17,18] Their magnetic susceptibilities exhibit an extended range of the Curie-Weiss (CW) behavior down to very low temperatures and high values of the frustration parameter $f$, defined as the ratio of the absolute value of the CW temperature $\Theta_{CW}$ to transition temperature $T_N$, i.e. $f = |\Theta_{CW}|/T_N$. For $FeSc_2S_4$ and $MnSc_2S_4$, frustration parameters of 900 and 11, respectively, were found. These compounds reveal broad spin-liquid ($MnSc_2S_4$) and spin-orbital liquid ($FeSc_2S_4$) regimes for the temperature range $T_N < T < |\Theta_{CW}|$ characterized by strongly enhanced dynamic spin and orbital correlations as evidenced by neutron scattering experiments.[19,20] In aluminum oxide spinels with the A-sites occupied by Fe and Co ions no long-range order was detected down to 2 K despite the presence of strong AFM exchange interactions as indicated by their large negative CW temperature $\Theta_{CW}$

~ -100 K. At low temperatures these spinels undergo short-range order. For example, the magnetic susceptibility of $CoAl_2O_4$ exhibits spin-glass-like behavior below a freezing temperature $T_f \sim 5$ K. The frustration parameter, $f = |\Theta_{CW}|/T_f \sim 22$, attributes $CoAl_2O_4$ to the family of strongly frustrated magnets. Recent neutron-diffraction studies[21] revealed a liquid-like structural factor of the magnetic intensities and suggested a spin-liquid ground state for this compound.

The experimental observations of strong frustration in magnetic A-site spinels are rather intriguing. Indeed, the A-site ions form a diamond lattice and assuming only nearest-neighbor interactions no frustration effects are expected. The A-site lattice can be viewed as consisting of two interpenetrating frustrated *fcc* sublattices shifted by (0.25, 0.25, 0.25) along the [111] direction. Considering second-neighbor exchange interactions that couple spins within each *fcc* sublattice, frustration effects can be anticipated, as was already noticed in Refs. 17 and 18 and has been explicitly demonstrated in recent theoretical studies.[22] The theoretical treatment of the exchange interactions in A-site antiferromagnetic spinels documented the crucial role of the second-neighbor AFM exchange for generating strong frustration effects and predicted a spiral spin-liquid ground state for a number of spinels with A-site magnetic ions.[22]

It is important to note that the finding of strong AFM exchange between the A-site $Co^{2+}$ ions in $CoAl_2O_4$ is unusual and at present is far from being fully understood. A relatively large distance (~ 3.5 Å) between nearest-neighbor $Co^{2+}$ ions excludes the direct exchange. Lotgering[23] and Roth[24] ascribed the large Curie-Weiss temperatures of the transition-metal aluminum oxide spinels to the inversion effect which can enhance the exchange between the A-site magnetic cations. Another explanation was proposed in Ref. 18, where the large $\Theta_{CW}$ was attributed to a high multiplicity of the exchange interaction paths between the tetrahedral ions, which include at least three intermediate ions.[17,24,25] The magnetic structure of Co-Al-O spinels is also far from being clarified. Earlier neutron-diffraction studies[24] failed to find clear long-range order in $CoAl_2O_4$ suggesting at the same time a collinear antiferromagnetic arrangement of the A-site spins of the nearest neighbor $Co^{2+}$ ions similar to that of $Co_3O_4$ as revealed by neutron diffraction.[26] However, the recent theoretical analysis of the spin order in A-site magnetic spinels[22] does not support such a simple spin arrangement.

The above mentioned problems motivated our interest in the $Co[Al_{1-x}Co_x]_2O_4$ system in which it is possible to vary the inversion degree and thus to study the effect of B-site substitution on the magnetic-ground state properties. Here we present the results of x-ray diffraction, magnetic susceptibility, electron-spin resonance, and heat capacity studies of $Co[Al_{1-x}Co_x]_2O_4$ spinels in the whole concentration range $0 \leq x \leq 1$ of substitution of Al by Co. We analyze the evolution of the magnetic correlations yielding a spin-glass ground state in frustrated $CoAl_2O_4$ and long-range antiferromagnetic order in $Co_3O_4$.

## II. EXPERIMENTAL DETAILS

Polycrystalline samples of $Co[Al_{1-x}Co_x]_2O_4$ with concentrations $x = 0$; 0.1; 0.2; 03; 0.4; 0.45; 0.5; 0.6; 0.75; 09, and 1.0, were prepared by solid-state reaction from high purity binary oxides CoO and $Al_2O_3$ (both 99.99%) and $Co_3O_4$ (99.985%). The synthesis was performed in alumina crucibles placed in evacuated quartz ampoules. In order to obtain minimal inversion which is essential for the low concentration range $x$, the sintering temperature was kept at 1000°C similar to the procedure reported in Ref. 16. After 3 days of sintering at the highest temperature the samples were slowly cooled to room temperature with a cooling rate of about 15°C/h. The synthesis was repeated several times in order to reach better homogeneity and full chemical reaction of starting materials. To reduce the oxygen non-stoichiometry the first synthesis was performed in an atmosphere of oxygen excess. Additional post-sintering annealing of several samples in flowing oxygen was also performed. The presence of non-reacted oxides in samples was checked by x-ray diffraction and EDX analysis.

Powder x-ray diffraction was performed by $Co-K_\alpha$ radiation with a wave length $\lambda = 1.78919$ Å utilizing X`Pert Pro MPD with X`Celerator (PW3040), Panalytycal. The data were analyzed by standard Rietveld refinement using the FULLPROF program.[27] Magnetization was measured with a commercial SQUID magnetometer (MPMS-5, Quantum Design), the heat capacity- using a Physical Properties Measurement System (PPMS, Quantum Design), both in the temperature range $2 K \leq T \leq 300 K$. The electron-spin resonance studies were carried out with a Bruker ELEXSYS E500 CW-spectrometer at X-band frequency ($\nu = 9.36$ GHz) in a helium gas flow cryostat (Oxford Instruments) for temperatures between 4.2 and 300K.

## III. EXPERIMENTAL RESULTS

### A. Structural data

X-ray powder diffraction experiments were performed at room temperature on powdered polycrystalline samples. The diffraction patterns showed single phase materials for all investigated compositions without any evidence of impurity phases (except for the sample with $x = 0.6$ where $\sim 2\%$ CoO were detected also by EDX analysis.). The x-ray data were analyzed assuming a cubic spinel structure with the space group $Fd\overline{3}m$ (No. 227). Within the Rietveld refinement, 12 parameters have been fitted: scale factor, zero point shift, resolution parameters, lattice constant, and oxygen positional parameter, occupation factor of cations and isotropic temperature factors for Co, Al and O ions. In the fitting procedure the occupation factor of oxygen was fixed to unity. The lattice constants $a_0$, the oxygen positional parameters $x_O$ and the occupation factors for the Co and Al cations on the A and B-sites are listed in Table 1.

Representative diffraction profiles for several samples with different Co concentrations together with the Rietveld refinement and the difference pattern are shown in Fig. 1. We note that the reflections (111) and (222) reveal a monotonous evolution in intensity with increasing $x$ and the reflection (533) shows a monotonous shift to high angles (see Fig. 2a and b). With increasing Co concentration on the B-sites the lattice constant $a_0$ is found to decrease. Our data for the both end members of the Co-Al-O system agree well with published results: $a_0 = 8.1063$ Å for $x = 0$ (Ref. 28) and $a_0 = 8.0837$ Å for $x = 1$ (Ref. 29). The concentration dependence of $a_0$ follows Vegard's law (see Fig. 2c). Deviations from a linear dependence can probably be attributed to deviation from the nominal Co concentration at the B-sites. Indeed, the refinement shows that for the substitution $x < 0.3$ the amount of Co on the B-sites is slightly higher than the nominal value. At the same time, a respective amount of Al ions is found on the tetrahedral A-sites. For substitutions $x > 0.3$ the amount of Co on the B-sites is close to the nominal value, except for the impure sample $x = 0.6$. The possible oxygen non-stoichiometry could also contribute to deviations from the Vegard's law; however, as grown samples and those additionally annealed in oxygen gave similar results and, therefore, this effect was difficult to estimate. Thus, the x-ray diffraction indicates on the formation of a complete set of solid solutions in this system and excludes coexistence of phases with different $x$.

### B. Magnetic susceptibility

Fig. 3a presents the temperature dependences of the inverse molar susceptibilities $\chi^{-1} = (M/H)^{-1}$ of $Co[Al_{1-x}Co_x]_2O_4$ measured on cooling in an external magnetic field of $H = 10$ kOe for temperatures $1.8$ K $\leq$ T $\leq 400$ K. Note that all samples exhibit rather similar values of the susceptibility for T $> 100$ K with a strictly linear temperature dependence of $\chi^{-1}(T)$. The inverse susceptibility $\chi^{-1}(T)$ of $CoAl_2O_4$ ($x = 0$) obeys a Curie-Weiss law for an extended temperature range above $10$ K. It shows an upturn at low temperatures which becomes more pronounced and shifts to higher temperature with increasing substitution. At the same time, the susceptibility starts to deviate from a CW law at higher temperatures indicating increasing spin correlations upon doping. For $Co_3O_4$ ($x = 1$) these deviations are clearly visible already below 100 K.

Table 1. Lattice constants $a_0$, oxygen positional parameter $x_O$ in fractional coordinates (f.c.) and occupation factors of Co and Al cations on the A- and B-sites of $Co[Al_{1-x}Co_x]_2O_4$ obtained by Rietveld analysis. R factors and $\chi^2$ values for each refinement are also indicated.

| Nominal concentra-tion , x | $a_0$ (Å) | $x_O$ (f.c.) | Co / A | Al / A | Al / B | Co / B | Rf | $\chi^2$ |
|---|---|---|---|---|---|---|---|---|
| 0 | 8.1070(1) | 0.2626 (3) | 0.896 (8) | 0.104 (8) | 0.936(6) | 0.064 (6) | 4.5 | 2.2 |
| 0.1 | 8.1029 (1) | 0.2631 (2) | 0.960 (4) | 0.040 (4) | 0.872 (2) | 0.128 (2) | 3.4 | 2.3 |
| 0.2 | 8.1007 (1) | 0.2636 (2) | 0.972 (4) | 0.028 (4) | 0.782 (2) | 0.218 (2) | 4.0 | 2.3 |
| 0.3 | 8.0980 (1) | 0.2639 (2) | 0.972 (4) | 0.028 (4) | 0.678 (2) | 0.322 (2) | 3.3 | 2.2 |
| 0.4 | 8.0973 (1) | 0.2640 (2) | 0.972 (4) | 0.028 (4) | 0.590 (2) | 0.410 (2) | 3.4 | 2.3 |
| 0.45 | 8.0961 (1) | 0.2639 (2) | 0.976 (4) | 0.024 (4) | 0.542 (2) | 0.458 (2) | 3.4 | 2.4 |
| 0.5 | 8.0962 (1) | 0.2639 (2) | 0.980 (4) | 0.020 (4) | 0.494 (2) | 0.506 (2) | 4.9 | 2.9 |
| 0.6* | 8.0928 (1) | 0.2620 (3) | 0.996 (4) | 0.004 (4) | 0.418 (2) | 0.582 (2) | 2.2 | 3.4 |
| 0.75 | 8.0891 (1) | 0.2642 (3) | 0.996 (4) | 0.004 (4) | 0.248 (2) | 0.752 (2) | 3.6 | 3.5 |
| 0.9 | 8.0864(1) | 0.2645(3) | 0.992(4) | 0.008(4) | 0.116(2) | 0.884(2) | 5.1 | 2.7 |
| 1 | 8.0832 (1) | 0.2656 (5) | 1.000(8) | - | - | 1.000(4) | 7.4 | 4.1 |

*Sample $x = 0.6$ contains ~ 2 % of CoO impurity.

The temperature dependences of the static susceptibility M/H in the low temperature range for 2 K ≤ T ≤ 50 K measured in a field of 1 kOe are shown in Fig. 3b. All samples with $x < 0.9$ exhibit a notable temperature hysteresis between zero-field cooled (ZFC) and field-cooled (FC) data. However, the susceptibility behavior is quite different for the samples with $x < 0.3$ and $x > 0.3$. For $x = 0$ the susceptibility shows a sharp anomaly at $T_f = 4.7$ K with a maximum in the ZFC and an upturn in FC susceptibility at temperatures below $T_f$, resembling spin-glass (SG) behavior. Above $T_f$, FC and ZFC susceptibilities coincide. The samples with $x = 0.1$ and $x = 0.2$ reveal SG-like behavior as well and freezing temperatures $T_f$ similar to $x = 0$. In addition, another broad maximum evolves above $T_f$ which becomes more pronounced with increasing $x$. For the sample with $x = 0.3$ the SG features are hardly discernible at 5 K and are absent for the samples with $x > 0.3$. With increasing substitution the broad maximum in the susceptibility is continuously shifted to higher temperatures. The susceptibility of the samples with $0.3 < x \leq 0.75$ still shows a ZFC-FC hysteresis but the splitting of the FC and ZFC curves occurs below the temperature of the susceptibility maximum. No hysteresis was detected for the samples with $x = 1$ and $0.9$.

To clarify the origin of the low-temperature irreversibilities we measured the *ac* susceptibility $\chi_{ac}$ for several samples in a frequency range 13 - 10000 Hz. Respective data for the sample with $x = 0.1$ are given in Fig. 4a. With increasing frequency $\nu$ the maximum of the real part of the *ac* susceptibility at $T_f$ shifts to higher temperatures, like in conventional spin glasses. The frequency shift of $T_f$, characterized by the quantity $dT_f/d(\log \nu)$, varies for the different frequencies within 0.04-0.06 and is close to values observed in spin glasses.[30] The *ac* susceptibility for the sample $x = 0.4$ behaves completely different. Like the static susceptibility $\chi_{DC}$, the *ac* susceptibility $\chi_{ac}$ shows a broad maximum with practically no frequency dependence (Fig. 4b).

The paramagnetic Curie-Weiss temperature $\Theta_{CW}$ and the effective magnetic moment $p_{eff}$ were determined from CW fits to the experimental susceptibility $\chi(T)$

$$\chi = \chi_{dia} + \frac{\mu_B^2}{3k_B}\frac{p_{eff}^2}{T - \Theta} \qquad (1)$$

The value of the temperature independent susceptibility of non-magnetic $ZnAl_2O_4$, $\chi_{dia} = -7.4 \cdot 10^{-5}$ emu/mol, taken from Refs. 18 and 31, was used to correct the experimental data for diamagnetic contributions.

Table 2 summarizes the fit parameters, $p_{eff}$ and $\Theta_{CW}$ obtained for different samples. Our experimentally determined Curie-Weiss temperatures and effective moments for the end members of this systems, $CoAl_2O_4$ ($x = 0$) and $Co_3O_4$ ($x = 1$), are in reasonably good agreement with those reported in earlier studies: $\Theta_{CW} = -104$ K, $p_{eff} = 4.65 \ \mu_B$ for $x = 0$ (Ref. 18) and $\Theta_{CW} = -109$ K, $p_{eff} = 4.78 \ \mu_B$ for $x = 1$ (Ref. 32). We note that the Curie-Weiss temperatures as well as the effective moments do not exhibit any systematic variation as a function of concentration. Surprisingly, all samples reveal similar $\Theta_{CW}$ values between -100 K and -120 K. The effective moment $p_{eff}$ for the different samples scatters within ~3 % that implies that the deviation from nominal stoichiometry of the samples is rather small. The deviations of the concentration of Co ions at the A-sites from the nominal value estimated by Rietveld refinement are of the same order. However with decreasing $x$ these deviations continuously increase while the effective moments show a non-systematic behavior. Therefore we cannot exclude some contribution to the variation of $p_{eff}$ from oxygen non-stoichiometry or CoO impurity with concentrations below the limit of detection (~2%). The effective g-factors $g_{eff}$ calculated from the susceptibility data, assuming that only $Co^{2+}$ ions with S = 3/2 contribute to the magnetic properties, range within 2.45 - 2.53 for the samples with different $x$. Similar values for $Co^{2+}$ are observed in different solids.[33] These values are however, by about 10% larger than the value of 2.23 obtained directly from the electron-spin resonance experiments (see paragraph C). To understand the reason of this inconsistency we performed post-sintering annealing of several samples in oxygen atmosphere. Annealing of the sample with $x = 0.5$ resulted in an effective g-factor of 2.26 in good agreement with the ESR data. However, for the other samples the annealing experiments were not so conclusive and the difference between these two g-factors remained. The most important fact is that the value of the Curie-Weiss temperature for the annealed samples did not change essentially, being at the same level of -100 K. This suggests that $\Theta_{CW}$ is only weakly influenced by the oxygen non-stoichiometry. Summarizing, we notice that these results as well as the similarity of our data and those obtained on the samples prepared at the lowest possible temperatures in Ref. 32, indicate that deviations from stoichiometry in our samples are indeed small. Therefore, we conclude that the non-stoichiometry and impurities have only marginal effect, and the Curie-Weiss temperature $\Theta_{CW} = -110 \pm 10$ K can be considered as an intrinsic value for all samples of the $Co[Al_{1-x}Co_x]_2O_4$ system.

Fig. 5 presents the magnetization curves measured at 2 K for samples with different $x$. For samples with $x \leq 0.3$ the magnetization curves show a small non-linearity (of about 2 % in the whole measured field range up to 50 kOe). With increasing $x$ the difference between the dc-susceptibility M/H at low and high fields becomes more pronounced. For example, for the sample with $x = 1$ a clear change from the low-field to high-field susceptibility occurs at a critical field of 9 kOe, resembling a spin-flop transition. An additional peculiarity is related to a linear reduction of the moment with increasing $x$, as shown in the inset of Fig. 5. The moment as measured at 50 kOe is reduced by 30 % when moving from $x = 0$ to $x = 1$.

TABLE 2. Curie-Weiss temperatures $\Theta_{CW}$ and effective moments $p_{eff} = g_{eff} [S(S+1)]^{1/2}$ determined from the CW fits to the magnetic susceptibility of $Co[Al_{1-x}Co_x]_2O_4$. Also given are Néel temperatures $T_N$, freezing temperatures $T_f$, temperature of the inflexion point of the susceptibility T*, and frustration parameters $f = |\Theta_{CW}|/T_f(T_N, T*)$.

| Substitution x | $\Theta_{CW}$ (K) | $p_{eff}$ ($\mu_B$) | $T_N/T_f$ (K) | T* (K) | f |
|---|---|---|---|---|---|
| 0 | -119 | 4.88 | - / 4.7 | - | 25.3 |
| 0.1 | -102 | 4.76 | - / 4.8 | - | 21.3 |
| 0.2 | -101 | 4.76 | - / 4.5 | - | 22.4 |
| 0.3 | -111 | 4.86 | - / 4.5 | 6.5 | 17.1 |
| 0.4 | -106 | 4.79 | | 6.8 | 15.6 |
| 0.45 | -101 | 4.75 | | 7.5 | 13.5 |
| 0.5 | -100 | 4.75 | | 7.0 | 14.3 |
| 0.6 | -100 | 4.81 | | 9.0 | 11.1 |
| 0.75 | -100 | 4.77 | | 11.5 | 8.7 |
| 0.9 | -110 | 4.78 | 21.6 / - | 20.8 | 5.3 |
| 1 | -114 | 4.80 | 29.5 / - | 29.2 | 3.9 |

## C. Electron-spin resonance

To gain insight into the local magnetic properties we measured the electron-spin resonance for samples throughout the complete substitution range. A typical ESR spectrum of the sample with $x = 0.9$ is shown in Fig. 6a. The ESR line can only be satisfactorily described assuming two lines with Lorentzian shape with different linewidth. Due to the large linewidth, the ESR signal was evaluated taking into account both circular components of each line as described by Ivanshin et al.[34] For the $x = 0.9$ sample, the intensity of the narrow line exhibits only a weak temperature dependence for 40 K < T < 300 K, while the intense broad line shows a pronounced temperature dependence (see Fig. 6b). In contrast, for pure $Co_3O_4$ both lines reveal comparable intensity and temperature dependences (see inset in Fig. 6b). The resulting intensity summed over both lines can be well fitted by a Curie-Weiss law with the CW temperature in good agreement with that obtained from the susceptibility data. With decreasing concentration $x$ the intensity of the broad line becomes dominating. The temperature dependences of the ESR linewidth $\Delta H$ for several samples with different substitution are shown in Fig. 7. For pure $Co_3O_4$ both broad and narrow lines exhibit the smallest linewidth (1600 and 740 Oe, respectively, at 100 K) and a smooth decrease on decreasing temperature. On approaching 30 K a sharp increase of $\Delta H$ is observed indicating the onset of strong local fields due to the development of AFM correlations. With decreasing concentration of the B-site Co ions a strong increase of the linewidth compared to that for $Co_3O_4$ occurs. For $x = 0.9$ the linewidth first shows a decrease on decreasing temperature similar to $Co_3O_4$ but below ~100 K $\Delta H$ rises again. Finally, on approaching 30 K, the linewidth for all substituted samples exhibits a strong increase, however less pronounced than for $Co_3O_4$. With further decreasing $x < 0.4$ the lines become so broad that they cannot be reasonably evaluated from X-band ESR measurements.

### D. Specific heat

The temperature dependences of the molar heat capacity C(T) for several $Co[Al_{1-x}Co_x]_2O_4$ samples with different $x$ are presented in Fig. 8a. The heat capacity for nonmagnetic $ZnAl_2O_4$ gives an estimate for the phonon contribution and is also included in this figure. The results for $Co_3O_4$ are in good agreement with those reported earlier by Khriplovich *et al.*[35] The heat capacity of this compound shows a sharp λ-type anomaly at 29.5 K unambiguously revealing a transition into a long-range ordered magnetic state. For $x = 0.9$ the λ-anomaly is shifted to lower temperature ($T_N = 21.6$ K), while a small peak can also be detected at 29 K. With further decreasing $x$, and a concomitant increase of the Al concentration on the B-site, the anomaly at ~29 K becomes less pronounced but still clearly discernible down to $x = 0.5$. It finally vanishes for $x \leq 0.45$. At low temperatures the specific heat for all samples with $x \leq 0.75$ develops an additional broad anomaly (hump) which is continuously shifted to lower temperatures on decreasing concentration for $0.3 < x \leq 0.75$.

We observed that for $x = 1$ and 0.9 the Néel temperature $T_N$ determined from the λ-anomaly in specific heat is very close (within 1K) to the temperature T* of the inflexion point of the susceptibility. At the same time, the values of $T_N$ and T* are about 10 K lower than the temperature of the broad maximum in the susceptibility, which is located at 39.7 K ($x = 1$) and 31.5 K ($x = 0.9$), respectively. For $x \leq 0.75$, the temperature of the inflexion point of the susceptibility T*comes close to the temperature $T_k$ of the hump in the specific heat. The values of $T_N$ determined from the specific heat and as well as of the freezing temperature $T_f$, and the temperature T* of the inflexion point of the susceptibility for different concentrations are given in Table 2.

The specific heat of the non-magnetic isostructural compound $ZnAl_2O_4$, corrected for the mass difference as suggested in Ref. 36, was used to estimate the phonon contribution in $Co[Al_{1-x}Co_x]_2O_4$. The magnetic part of the specific heat $C_m$ was calculated by subtraction of the respective phonon contribution from the total specific heat. The temperature dependence of the change of magnetic entropy calculated by integration of the magnetic part of the specific heat over the measured temperature range for several samples is shown in Fig. 8*b*. For $x = 1$ the magnetic entropy at the Néel temperature reaches about 50 % of the full magnetic entropy $Rln(2S+1)$ expected for the ordering of the $Co^{2+}$ ions with spin $S = 3/2$. This value of the magnetic entropy at $T_N$ in $Co_3O_4$ is comparable to that observed in conventional antiferromagnets. We also note a shift of the magnetic entropy to lower temperatures with decreasing $x$. For $x = 0.9$ the magnetic entropy at $T_N$ reaches only ~38 % of the full value, whereas for samples with $x \leq 0.2$ the magnetic entropy at the freezing temperature is 7 %, only.

Finally, we want to mention an important observation in the behavior of the specific heat for temperatures below $T_k$ down to lowest temperatures (see inset in Fig. 8b). The magnetic part of the specific heat shows a change from a $T^2$ dependence for the sample with $x = 0$ to a $T^{2.5}$ dependence for the samples with $x$ between 0.3 and 0.75, and finally, to ~$T^3$ temperature dependence for $x = 0.9$ and 1. We are aware that the temperature range of this observation is less than one decade and measurements to lower temperatures are necessary to arrive at final conclusions. But we think that this observation is too significant to be ignored.

## IV. DISCUSSION

The first important result of our study is related to the high values of the Curie-Weiss temperatures observed in Co-Al spinels. We would like to emphasize that the observation of very similar CW temperatures of the order of -110 K for compositions with different Co concentrations on the B-sites apparently excludes the inversion mechanism suggested earlier as reason for the enhanced CW temperature of $CoAl_2O_4$ and of the other magnetic A-site aluminum-oxide spinels.[23,25] The alternative explanation[18] based on the high multiplicity of the superexchange interactions between the A-site $Co^{2+}$ ions seems to be more plausible. However, the fact that the magnetic exchange in our system is equally strong for both B-site $Co^{3+}$ ions in a low-spin $3d^6$ state and diamagnetic $Al^{3+}$ ions is surprising. At one hand, the independence of the CW temperatures on substitution suggests that the exchange between tetrahedral $Co^{2+}$ ions is almost independent from the ions at the octahedral sites. On the other hand, the appearance of a broad maximum in the susceptibility that shifts to higher temperatures with increasing $x$, indicates increasing AFM spin correlations. Finally, the establishment of a true long-range antiferromagnetic order for $x = 0.9$ and 1.0 has to be attributed to an increasing role of the exchange interactions through the B-site Co ions. This exchange appears to be relatively small as compared, for example, to $Co^{2+}$- O - $Co^{2+}$ exchange, but its contribution becomes evidently important at large $x$. The complexity of the magnetic phase diagram, depending on the ratio of the nearest neighbor ($J_1$) and next-nearest neighbor ($J_2$) exchange, has been modeled in Ref. 22. According to these calculations, $CoAl_2O_4$ with $J_2/J_1 \approx 1/8$ is just in the spiral spin liquid regime with large frustration, while with decreasing $J_2$ the Neel state is stabilized, which obviously happens for $Co_3O_4$.

The second important result is related to strong spin frustration which is evidenced by magnetic and specific heat measurements. For the samples with $x \leq 0.2$, an extended temperature range of the Curie-Weiss behavior of the static susceptibility, low freezing temperatures of ~5 K and large negative Curie-Weiss temperatures of the order of -110 K result in frustration parameters $f > 20$ (Table 2). In addition, the value of the magnetic entropy at the SG transition reaches only 7 % of the full magnetic entropy. This implies that a significant fraction of the magnetic entropy is released at high temperatures, which is a clear signature of frustration.

Concerning the SG-like phase, it is important to note the following: Although the static and $ac$ susceptibilities of the samples with low concentration, $x \leq 0.2$, show temperature and frequency variations resembling that of canonical spin glasses, a $T^2$-dependence of the specific heat observed for $CoAl_2O_4$ ($x = 0$) and roughly the same for the sample with $x = 0.1$ is in sharp contrast with that expected for spin glasses. For canonical spin glasses a linear temperature dependence of the specific heat is observed below the cusp in C(T) which appears above the freezing temperature.[30] We also note that a $T^2$-dependence of the specific heat was already found in a number of geometrically frustrated spin glasses with spinel and Kagomé structure[37,38] and in the strongly frustrated spinel $FeSc_2S_4$ with a spin-orbital liquid ground state.[19] Because the spin-glass behavior is not seen for $x \geq 0.3$, but is most pronounced in samples with lower $x$, its appearance can be attributed to disorder on the A-sites. Indeed, for the sample with the nominal concentration $x = 0$, which shows the largest SG irreversibility, a significant fraction of Al ions (of about 10 %) is found on the A-sites. For samples with larger $x$ the SG features become weaker which clearly correlates with the decrease of the disorder on the A-sites (see Table 1). Therefore, we think that the A-site disorder might be responsible for the release of spin frustration at lowest temperatures. In the absence of disorder on the A-sites in the sample with $x = 0$ one would probably expect a quantum paramagnetic behavior with a strict CW dependence of the susceptibility towards zero temperature. Preparation of purely stoichiometric $CoAl_2O_4$ is, however, a challenging problem, since in solid-state synthesis this compound forms only above 850 ${}^{o}C$, a temperature which is high enough to generate site intermixing.[39,40]

In contrast to compositions with SG-like ground state, the $x = 1$ end member of this system, $Co_3O_4$, develops a long-range antiferromagnetic state below $T_N = 29.5$ K, as revealed by the sharp anomaly in the specific heat and by the divergent behavior of the linewidth in the ESR experiments. Our magnetization data suggest that the spin configuration in this compound is more complex than the collinear antiferromagnetic arrangement proposed earlier in Ref 26. In fact, the observed pronounced non-linearity of the magnetization curve and the increase of the susceptibility at high fields above the critical field of 9 kOe (Fig. 5) indicate a field induced magnetic transition. Similar magnetization features have been observed in several spinels with incommensurate helical spin arrangement.[41] Helical spin order in these compounds is a consequence of the increasing contribution of next-nearest neighbor exchange interactions which compete with the nearest-neighbor interactions. The next-nearest neighbor interactions are particularly important in Co-Al-O spinels; therefore they also might be responsible for the complex arrangement of the A-site spins. After original submission of our paper, Ikedo et al.[42] reported on the µSR studies of $Co_3O_4$ which were interpreted in terms of an incommensurate spin configuration. By analogy with the strongly frustrated $MnSc_2S_4$ spinel[20] they attributed the incommensurate spin order in $Co_3O_4$ to a structural transition due to a charge- and/or spin-state change of the Co ions. However, recent studies of strongly frustrated magnetic chalcogenide spinels $ZnCr_2S(Se)_4$ obviously demonstrate that competing FM and AFM exchange interactions alone can drive the structural distortions and a structural phase transition.[12-15]

In Co-Al-O spinels a contribution to structural distortions from the spin-orbital coupling can be additionally anticipated from the ESR data. According to Ref. 43, the dominant contribution to the linewidth of the ESR spectra in $Co_3O_4$ arises from the splitting of the ground state spin quartet of $Co^{2+}$ ions in a uniaxial crystal field. As we deal with polycrystalline samples, we cannot resolve the corresponding anisotropy of the spectra, but only approximate the powder pattern by means of two Lorentzians due to the extreme lines which confine the angular distribution. The anisotropy and linewidth depend on temperature, hence the intensity weight may shift between the extreme lines and, thus, only the sum of both lines represents the real susceptibility, as has been shown by the fits of the intensity in Fig. 6. Substitution with Al results in increasing local distortions of the tetrahedral ligand field and the two lines existing in the ESR spectra imply both the increasing anisotropy as well as a wide distribution of the local distortions.

In the intermediate range $0.3 < x \leq 0.75$, both the susceptibility and specific heat show no pronounced anomalies characteristic for a long-range magnetic or SG-like ordering. Lack of a clear magnetic ordering in connection with high CW temperatures again points to the presence of strong spin frustration. A $T^{2.5}$ temperature dependence of the specific heat for these samples observed at the lowest temperatures suggests a spin-liquid (SL) ground state in accord with the theoretical predictions for the frustrated magnetic A-site spinels.[22] Preliminary neutron-scattering investigations[44] of the sample with $x = 0.35$, however, revealed a long-range AFM order, which sets in below the temperature of the inflexion point in the susceptibility T* coexisting with a liquid-like structure factor. At the same time, the intensity of the magnetic reflections was much lower than that for $Co_3O_4$ with pure long-range AFM ground state, suggesting that AFM ordering in $x = 0.35$ is not complete and possibly coexists with the spin-liquid phase.

The evolution of the magnetic states in $Co[Al_{1-x}Co_x]_2O_4$ depending on the B-site Co concentration and temperature that follows from our studies, can be interpreted within the phase diagram shown in Fig. 9. The spin-liquid regime is separated from the paramagnetic (PM) regime by a line corresponding to a temperature $|\Theta_{CW}|$ as suggested by recent experimental[19,20] and theoretical[22] investigations of the A-site frustrated magnets. However, as most of our experimental techniques are not sensitive to dynamic spin correlations, we are unable to verify this in detail. From the ESR data the appearance of such correlations below 100 K can be deduced from the line broadening for the samples with $x = 0.9$ and 0.45. For the other samples the ESR results were not so conclusive. Therefore, we use this well established relation as an upper limit for the boundary between PM and SL phases. It is clear that this is only an approximate boundary, and does not separate different phases. In the range $0.3 < x < 0.75$ the AFM order gradually develops with increasing $x$, probably coexisting with the spin-liquid state. It has to be noted that the specific heat of the samples with $0.5 \leq x \leq 0.75$ still exhibits a small but clearly discernible anomaly at ~29 K which is close to the Néel temperature of $Co_3O_4$. The amplitude of this anomaly at ~29 K shows a non-monotonous behavior with substitution. It is noticeable even in the sample with $x = 0.9$ which has a clear AFM ground state and a Néel temperature of 21.6 K. At the same time, no anomaly at 29 K is present in the samples with concentrations $0.3 \leq x \leq 0.45$, but they show a very similar behavior of the susceptibility and specific heat at low temperatures like samples with $0.5 \leq x \leq 0.75$. These facts suggest the absence of a true long-range AFM order at 29 K also for $0.5 \leq x \leq 0.75$. In order to understand the origin of the anomaly at 29 K we have simulated the behavior of the specific heat assuming a mixture of two phases: $Co_3O_4$ and $Co[Al_{1-x}Co_x]_2O_4$ with respective $x$. It was possible to describe the behavior of specific heat in the vicinity of 29 K assuming the presence of about 17 and 10 mol % of $Co_3O_4$ for $x = 0.75$ and 0.6, respectively. However such a large amount of the second phase of $Co_3O_4$ must be evidently detected by x-ray diffraction and ESR, which is in contradiction with the experimental data. Therefore, we think that the anomaly in the specific heat at 29 K observed in the samples with $0.5 \leq x \leq 0.9$ is not connected with an impurity $Co_3O_4$ phase. We suppose that it can be related to quenched disorder and built-in non-percolating clusters of $Co_3O_4$ statistically distributed in the volume of the samples which can yield a smeared anomaly in the specific heat at 29 K. The ESR data for these samples provide additional arguments for the intrinsic character of this effect. Indeed, the linewidth of the ESR signal for the sample $x = 0.45$ shows pronounced features at 29 K although no anomalies are present in the specific heat at this temperature. This indicates that, even if the heat capacity does not show any anomaly due to the reduced number of clusters, their effect on the whole spin system is still dominating the spin relaxation. We speculate on some similarities of the magnetic behavior in our Co-Al-O spinels and Griffiths phase singularities observed recently in disordered manganites.[45] But additional investigations are necessary to clarify the origin of this anomaly.

The left bottom part of the phase diagram refers to samples with $x \leq 0.2$ which exhibit spin-glass like behavior. However, as mentioned above, their ground state cannot be ascribed to a pure spin glass. The presence of strong spin frustration evidenced by the magnetic studies suggests a spin-liquid state also for these compositions in agreement with recent neutron-diffraction studies of $CoAl_2O_4$.[21] Theoretically, the origin of the $T^2$ dependence in geometrically frustrated magnets with a SG ground state is not understood. Several mechanisms including thermal and quantum fluctuations were proposed.[46] However, as we observed SG features and $T^2$ dependence in the samples with strong disorder on the A-sites, we suppose that the deviations of the specific heat at low temperatures from the theoretically predicted $T^{2.5}$ dependence for the spiral-spin liquid might be associated with the magnetic inhomogeneity due to structural disorder. A possible influence of magnetic disorder was already discussed for geometrically frustrated spin glasses which manifest a $T^2$-dependence of the specific heat.[37,38]

We would like to make a final remark related to the mechanism of spin frustration. In Ref. 18, where only a sample with $x = 0$ was studied, it was ascribed to geometrical frustration of the $Co^{2+}$ ions residing on the A-site tetrahedral network. Taking into account recent theoretical results concerning frustration in the magnetic A-site spinels[22] we reconsider the mechanism of spin frustration in Co-Al-O spinels and assign it rather to competing exchange interactions, i.e., to bond frustration.

## V. CONCLUSIONS

X-ray diffraction, magnetic susceptibility, electron-spin resonance, and specific heat investigations have been performed on the $Co[Al_{1-x}Co_x]_2O_4$ spinels in the range of $Co^{3+}$ concentrations $0 \leq x \leq 1$ on the octahedral B-sites. In the complete concentration range the samples reveal a large negative Curie-Weiss temperature of the order of -110 K. They are attributed to the high multiplicity of the superexchange interactions between the A-site cations, whereas the inversion mechanism is considered to be of minor importance. The specific heat of the samples with $x = 1$ and 0.9 manifests a pronounced $\lambda$-like anomaly at 29.5 K and 21.2 K, respectively, indicating a transition into the long-range ordered antiferromagnetic state. The Néel temperature is shifted to lower temperatures on decreasing $x$: it decreases sharply for $0.75 \leq x \leq 1.0$ and more smoothly for $x < 0.75$. The reduced temperature of the magnetic ordering indicates strong frustration, and the $T^{2.5}$ dependence of the specific heat suggests a spin-liquid ground state. The samples with $x \leq 0.2$ exhibit a spin-glass like behavior below 5 K revealing a high value of the frustration parameter $f > 20$. At the lowest temperatures the specific heat of these samples follows a $T^2$ dependence which is ascribed to structural disorder on the A-sites. The frustration mechanism is related to bond frustration due to strong competition of the exchange interaction between the A-site spins of $Co^{2+}$ ions.

## ACKNOWLEDGEMENTS

The authors are thankful to Mrs. B. Opitz for x-ray measurements and D. Vieweg for additional susceptibility measurements. Funding by Sächsisches Staatsministerium für Wissenschaft und Kunst under grant 4-7531.50-04-823-05/3 is gratefully acknowledged. The support of US CRDF and MRDA via BGP III grant MP2-3050 is also gratefully acknowledged. The experiments at the University of Augsburg were partially supported by the Deutsche Forschungsgemeinschaft within SFB 484 (Augsburg).

their samples show much stronger deviation from the CW law already at 50 K compared to our samples (10 K) and higher $T_k$ of 10 K in the specific heat than ours (7.6 K). We also note that our sample with $x = 0.3$ shows a similar behavior like their sample with $x = 0$.

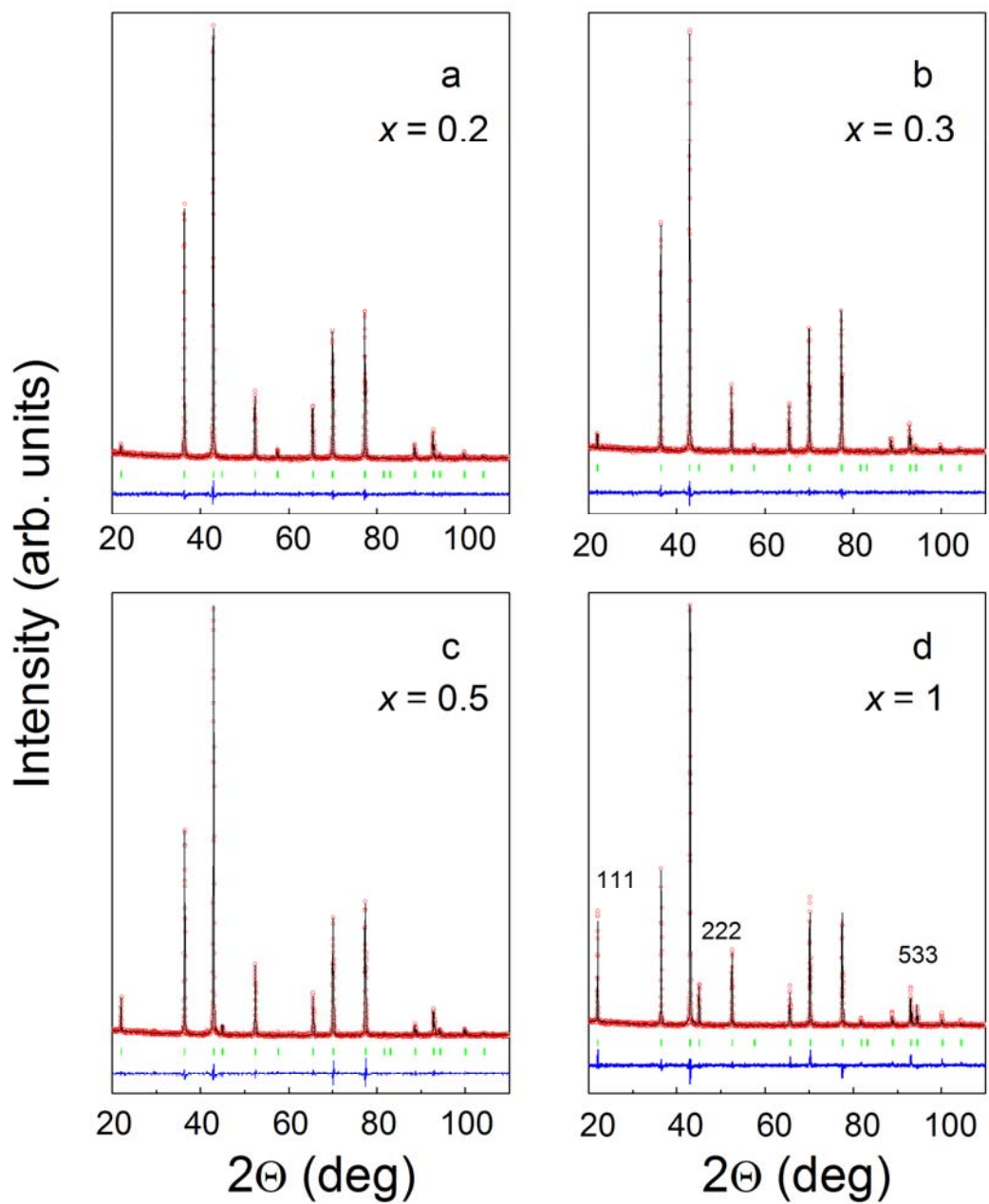

FIG. 1. (color online): x-ray diffraction profiles of Co[Al$_{1-x}$Co$_x$]$_2$O$_4$ with different Co concentrations $x$ at the B-sites. The measured intensities (red open circles) are compared with the calculated profile using Rietveld refinement (black solid line). The Bragg positions of the normal cubic spinel structure are indicated by vertical (green) bars and the difference pattern is shown by the bottom thin (blue) solid line.

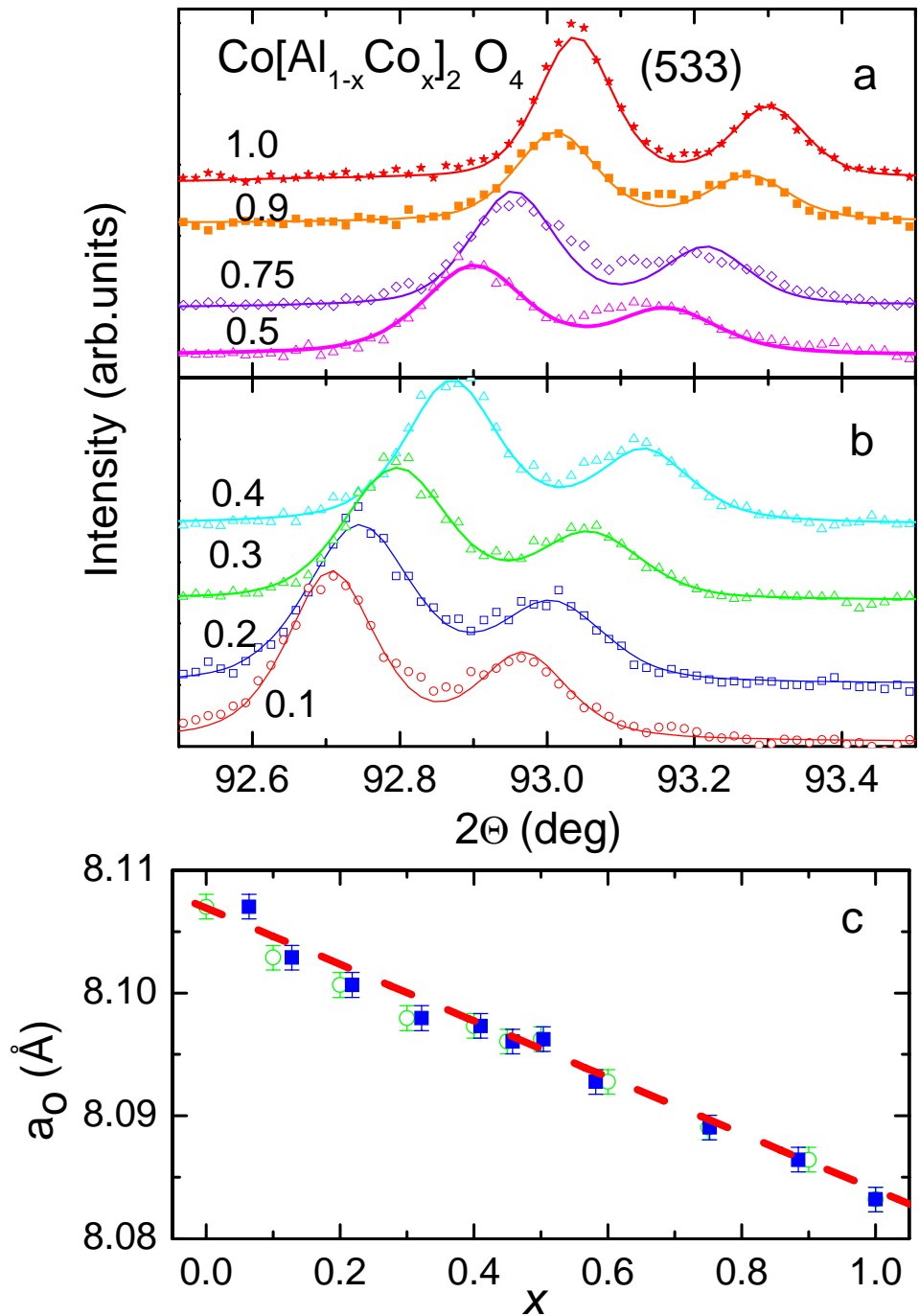

FIG. 2. (color online): a, b: evolution with substitution $x$ of the characteristic reflection (533); c: dependence of the lattice constant $a_0$ on $x$. Open symbols are related to nominal concentrations, closed symbols - to refined values of $x$.

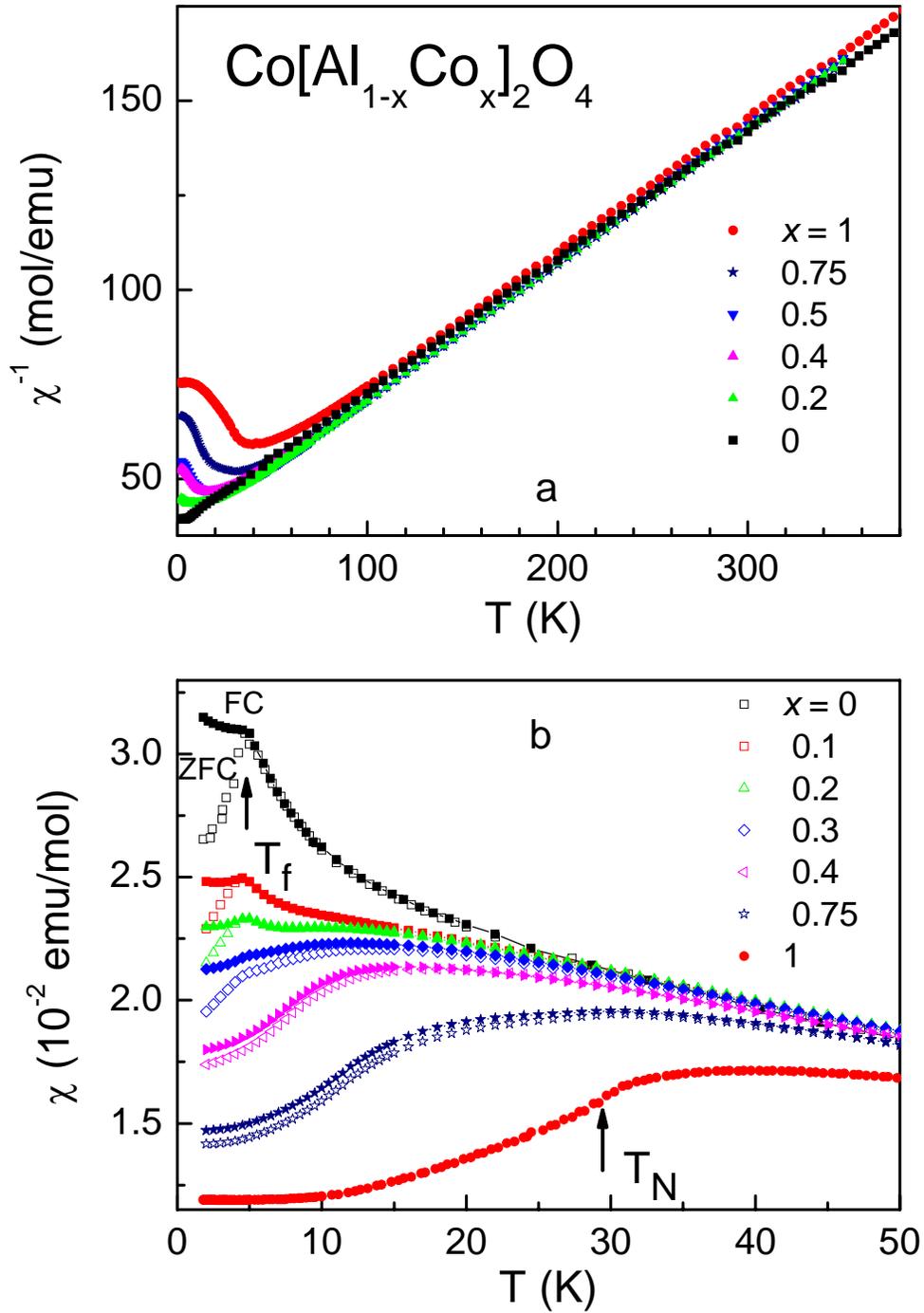

FIG. 3. (color online): a: Inverse molar susceptibility of Co[Al$_{1-x}$Co$_x$]$_2$O$_4$ vs. temperature in a field of 10 kOe for different concentrations *x*. b: Susceptibility vs. temperature in a field of 1 kOe. The arrows indicate the temperature of the ZFC susceptibility maximum, T$_f$, for *x* < 0.3 and the Néel temperature T$_N$ for *x* = 1.

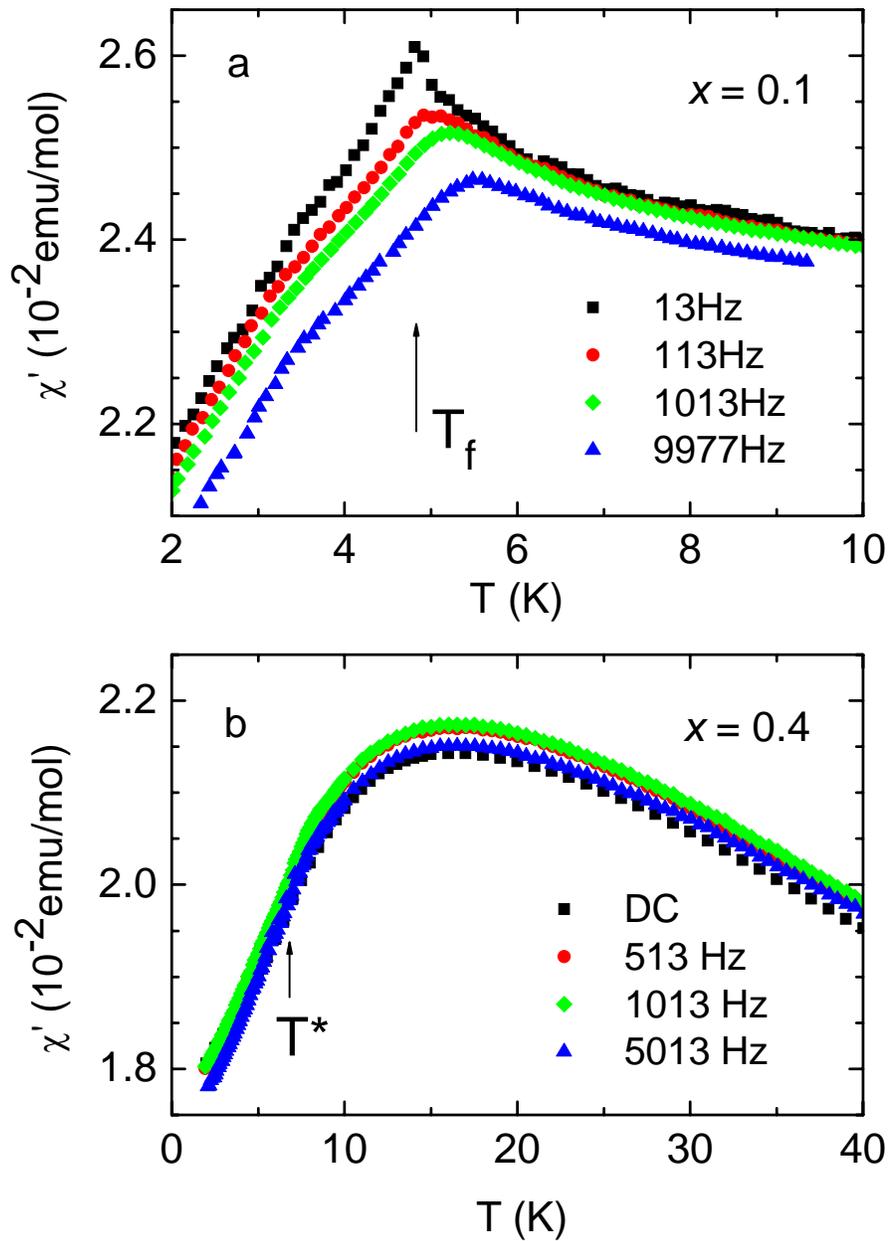

FIG. 4. (color online): Temperature dependences of the real part of the *ac* susceptibility for different frequencies for the samples: a) $x = 0.1$, b) $x = 0.4$. The static (DC) susceptibility is also shown. The arrows mark the freezing temperature $T_f$ and the temperature of the inflexion point of the susceptibility $T^*$.

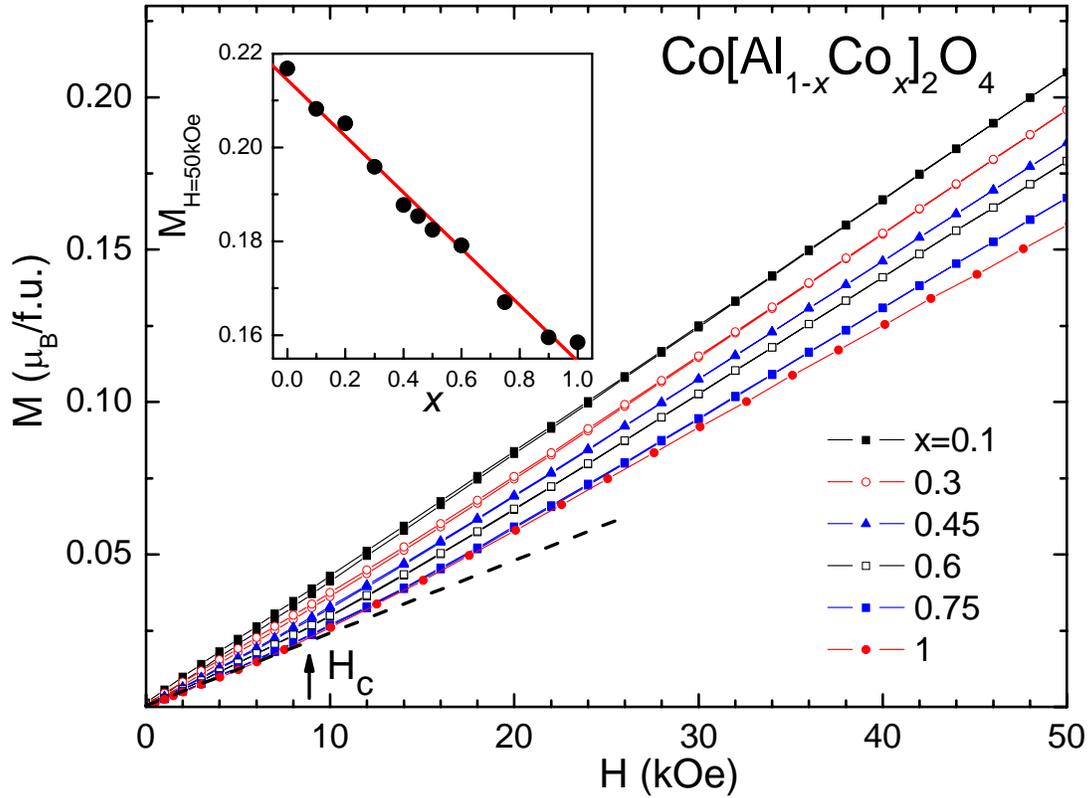

FIG. 5. (color online): Magnetization M(H) for Co[Al$_{1-x}$Co$_x$]$_2$O$_4$ samples with different *x* at T = 2 K. The arrow indicates a metamagnetic transition at a critical field H$_c$ ≈ 9 kOe for Co$_3$O$_4$. The dashed line traces the magnetization in fields below H$_c$ for the sample *x* = 1. Inset: Scaling of M in a field of 50 kOe with the substitution *x*.

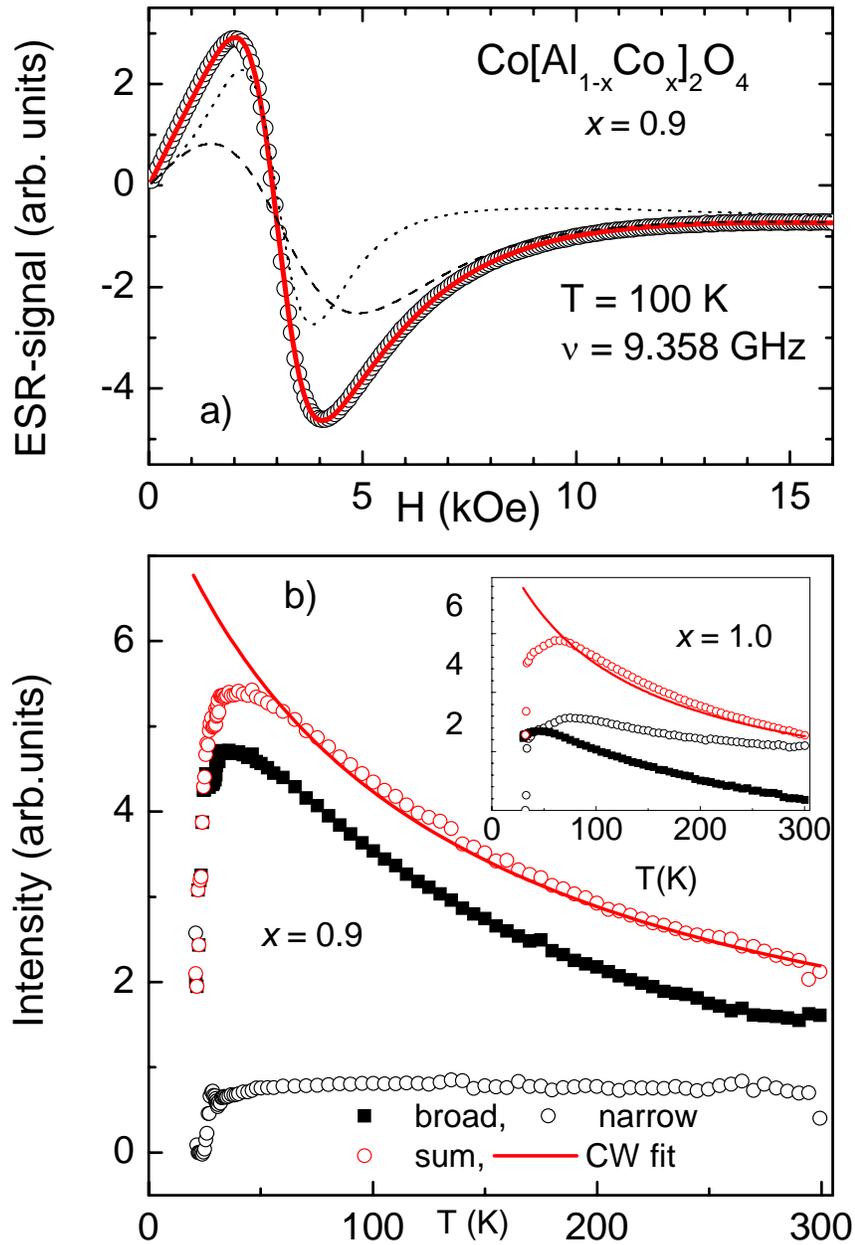

FIG. 6 (color online): a) ESR spectrum for the sample with $x = 0.9$ at 100 K. Circles represent experimental data, dashed, dotted and solid lines refer to deconvoluted broad line, narrow line and sum of both, respectively. b) Temperature dependence of the intensity of the broad, narrow and sum of both ESR lines for $x = 0.9$. Inset: the same for the sample $Co_3O_4$. Solid lines present the CW fit with $\Theta_{CW} = -113.5$ K and $-105$ K for $x = 0.9$ and 1.0, respectively.

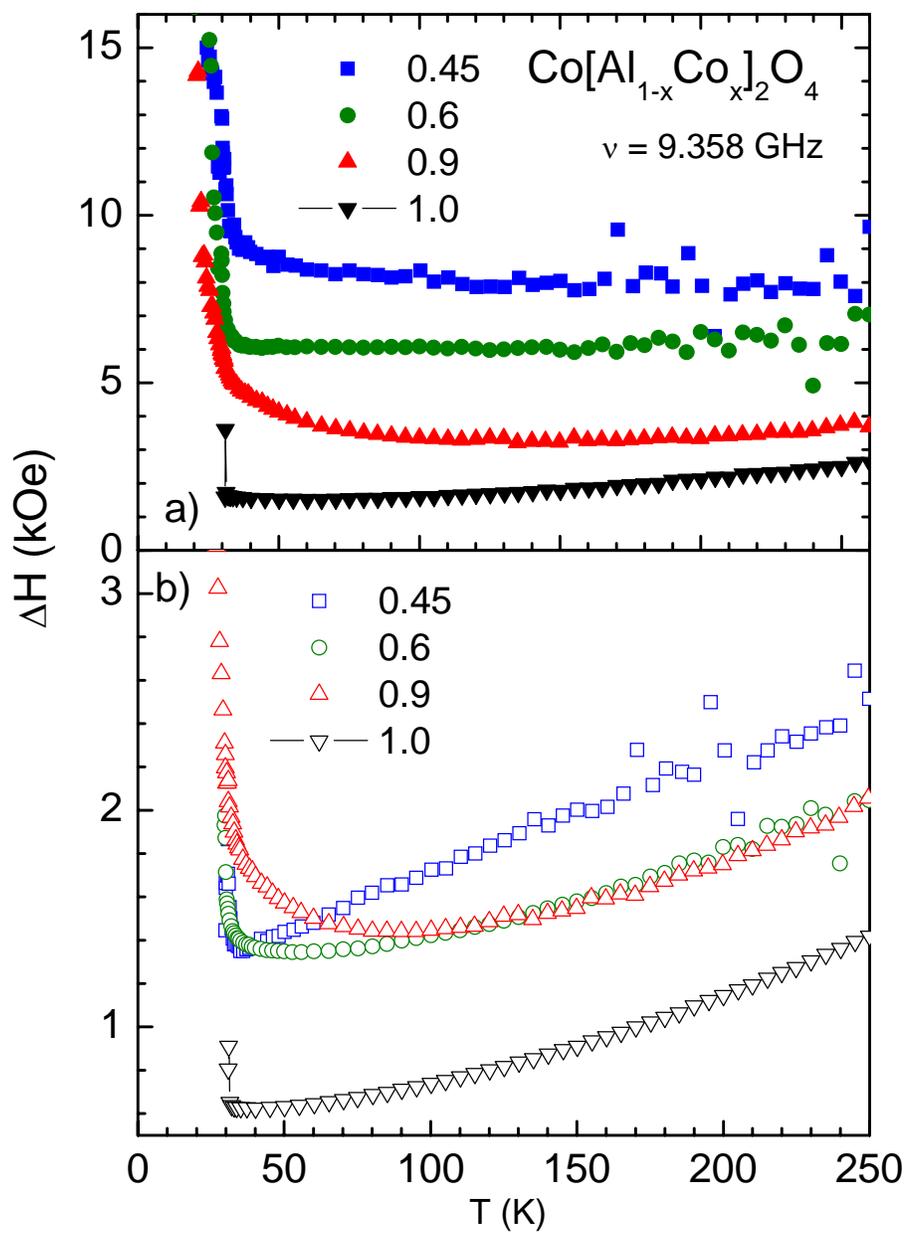

FIG. 7 (color online) Temperature dependence of the linewidth of the broad (a) and narrow (b) lines for samples with various concentrations *x*.

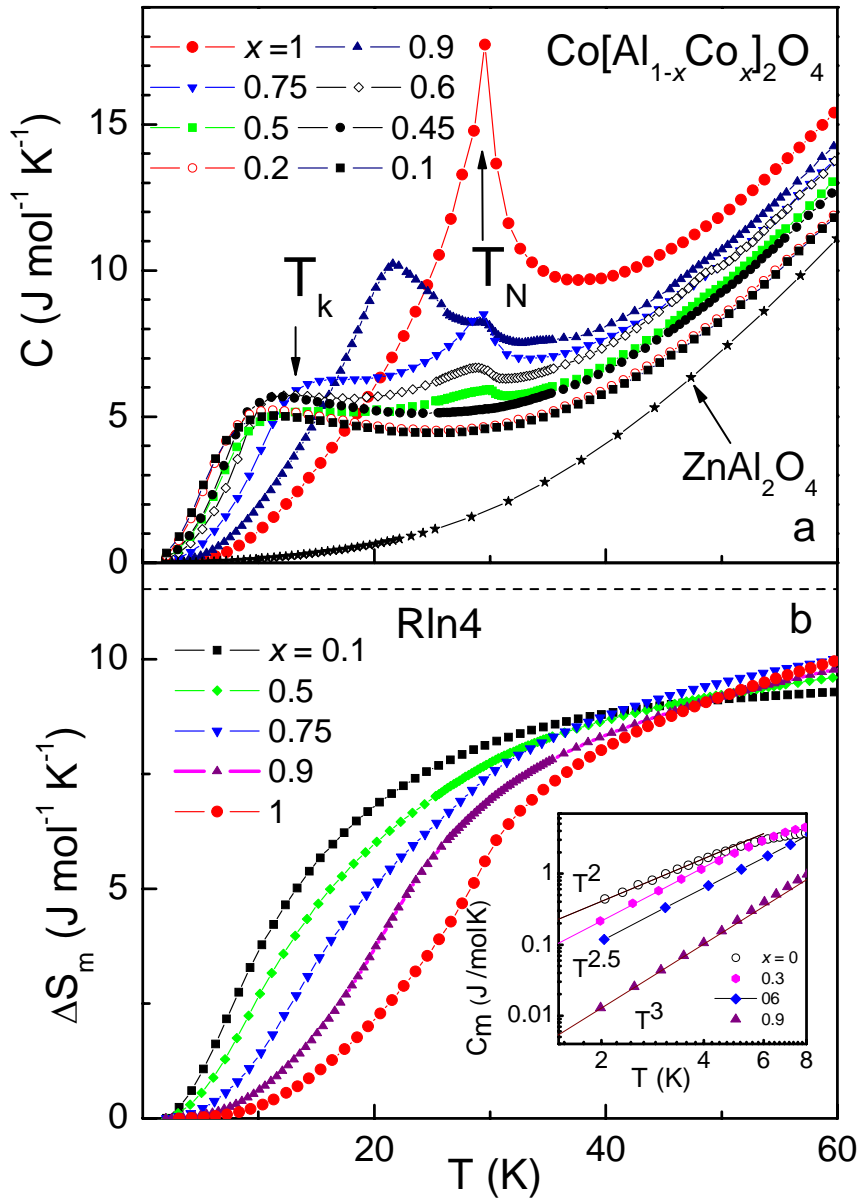

FIG. 8. (color online): a: Heat capacitiy C(T) of $Co[Al_{1-x}Co_x]_2O_4$ for different $x$. Arrows indicate the anomaly at the Néel temperature $T_N$ for $x = 1.0$ and of the kink at low temperatures at $T_k$ for $x = 0.75$. The lowest curve shows the heat capacity for non-magnetic $ZnAl_2O_4$; b) Change of the magnetic entropy $\Delta S_m(T)$. The dashed line shows the entropy corresponding to a value $Rln(2S+1)$ for $S=3/2$. Inset: Magnetic part of the specific heat $C_m$ at low temperatures for several compositions of $Co[Al_{1-x}Co_x]_2O_4$ on double logarithmic scale. The solid lines through the experimental points correspond to a $T^2$ dependence for $x = 0$, to a $T^{2.5}$ dependence for 0.3 and 0.6 and to a $T^3$ dependence for $x = 0.9$.

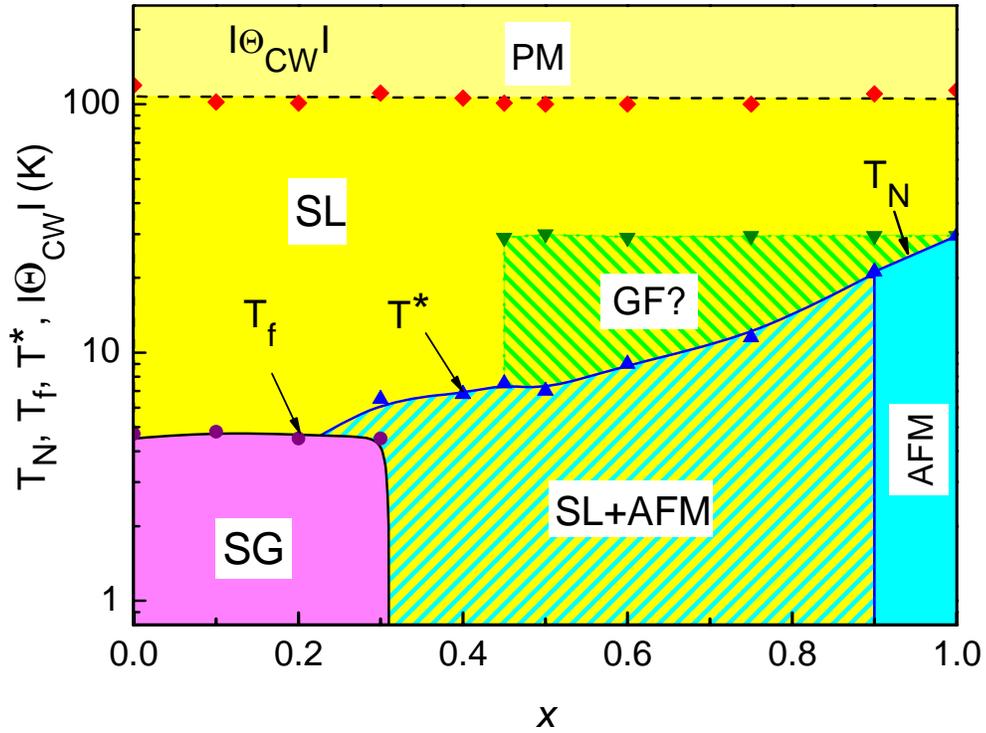

FIG. 9. (color online): Magnetic phase diagram of Co[Al$_{1-x}$Co$_x$]$_2$O$_4$. For $x = 0.9$ and $1.0$ the ground state is antiferromagnetic (AFM). The spin-glass (SG) state exists for $0 \leq x < 0.3$. For $0.3 \leq x \leq 0.75$ spin liquid (SL) and AFM phases coexist. The SL is separated from the paramagnetic (PM) state by a line corresponding to the Curie-Weiss temperature $|\Theta_{CW}|$. $T_N$, $T_f$, and $T^*$ define, respectively, Néel temperature, freezing temperature, and temperature of the inflexion point of the static susceptibility. For GF phase see text.